# First-principles discovery of stable, anisotropic, semiconducting $Sb_2X_2O$ (X = S, Se) and Janus $Sb_2SSeO$ nanosheets for optoelectronics and photocatalysis

Masoud Shahrokhi[a]* and Bohayra Mortazavi[b,c]

[a]Department of Physics, Faculty of Science, University of Ostrava, 30. dubna 22, 701 03 Ostrava, Czech Republic.

[b]Institute of Photonics, Department of Mathematics and Physics, Leibniz Universität Hannover, Welfengarten 1A, Hannover 30167, Germany.

[c]Cluster of Excellence PhoenixD, Leibniz Universität Hannover, Welfengarten 1A, Hannover 30167, Germany.

**ABSTRACT**

In this work, we conduct a comprehensive first-principles investigation into the design and discovery of novel antimony oxychalcogenide monolayers $Sb_2X_2O$ (X = S, Se) and Janus $Sb_2SSeO$, examining their structural stability, elastic, electronic, optoelectronic, and photocatalytic properties. Our analysis confirms their thermodynamic and dynamical stability and reveals low cleavage energies, indicating strong feasibility for mechanical exfoliation. The excellent agreement between our HSE06-predicted bandgap of bulk $Sb_2S_2O$ and experimental measurements further validates the employed computational framework. Electronic structure calculations including spin–orbit coupling show that the $Sb_2S_2O$ monolayer is a direct-gap semiconductor (~2.80 eV), while $Sb_2Se_2O$ and $Sb_2SSeO$ exhibit indirect band gaps of 2.24 and 2.44 eV, accompanied by pronounced anisotropy in effective masses and carrier mobilities. We also find that their optoelectronic responses can be efficiently tuned via biaxial strain, providing a viable route for device-specific property engineering. Favorable band alignments, strong optical absorption, efficient carrier transport, and relatively high dielectric constants collectively support their candidacy for overall water splitting under neutral conditions. These results establish a solid theoretical foundation for the rational design of Sb-based 2D nanostructures and highlight their potential in next-generation direction-dependent optoelectronic and sustainable energy-conversion applications.

**INTRODUCTION**

The urgent demand for sustainable energy solutions has accelerated the search for new semiconducting materials capable of driving photocatalytic and optoelectronic processes with high efficiency. Two-dimensional (2D) materials are at the forefront of this search due to their unique electronic structures, high surface-to-volume ratios, robust flexibility and stability, and tunable properties. The emergence of 2D materials dates back to 2004, when graphene was first isolated [1]. Since then, research and innovations in graphene have made fundamental contributions to condensed matter physics and materials engineering. Graphene's exceptional characteristics, including ultra-high carrier mobility [2] and superior mechanical strength [3], have inspired worldwide endeavors to discover other atom-thick materials. However, the semimetallic nature of pristine graphene, characterized by a zero electronic band gap at the Dirac cone, severely limits its direct applicability in mainstream electronics, optoelectronics, and energy conversion devices that fundamentally require intrinsic semiconductors with suitable band gaps. This critical requirement has necessitated a major shift in research focus toward designing, predicting, and synthesizing 2D materials with inherent semiconducting properties. As a result, a wide variety of 2D semiconductors have been synthesized to date, including transition metal dichalcogenides, like $MoS_2$ and $WS_2$ [4,5], $MoSi_2N_4$ and $WSi_2N_4$ [6], single-triazine-based $g-C_3N_4$ [7], Indium selenide InSe [8], penta-PdPS [9] and penta-PdPSe [10] and nanosheets. Although these materials provide band gaps that overcome the limitations of graphene, the continuous drive for improved performance in specific applications, such as faster transistors, high-efficiency solar cells, and robust photocatalysts, requires exploring new chemical compositions and structural motifs that offer superior stability and tailored electronic properties.

Among the promising new classes of 2D materials, those based on Group V elements, such as Antimony (Sb), have garnered significant attention [11,12]. Antimony-based compounds, like antimonene and $Sb_2X_3$ derivatives, often exhibit layered structures, high intrinsic stability, and high carrier mobilities, making them excellent candidates for exfoliation and integration into flexible nanodevices [13,14]. Their electronic band structures are highly tunable through chemical doping, strain engineering, or the creation of hybrid structures, offering a rich platform for developing next-generation functional materials. A particularly powerful strategy for customizing the properties of these nanosheets is the creation of Janus structures, meaning one face of the 2D sheet is terminated by a different element than the other (e.g., $Sb_2X_2$ becomes $Sb_2SSe$). This broken out-of-plane symmetry results in the generation of an intrinsic electric dipole moment, which modifies the electronic band alignment and dictates the flow of charge carriers perpendicular to the plane of the material. The dipole effect inherent to Janus structures is especially impactful in photocatalysis, which is a critical avenue for sustainable hydrogen production via water splitting. The internal electric field generated by the Janus configuration actively and directionally separates these photocarriers, dramatically suppressing the highly detrimental recombination process and thereby significantly boosting the overall photocatalytic quantum efficiency [15].

In the ongoing endeavors of predicting and experimentally realizing 2D semiconductors, the successful fabrication of layered $Sb_2S_2O$ structures was most recently demonstrated by Radatovic and colleagues [16], and the potential for device applications due to tunable structural, chemical, and optical properties was highlighted. Their measurements confirmed a bandgap of 2.04 eV, high photoresponsivity, ambient stability without protective coatings, and promising prospects for polarization-sensitive electronic and optoelectronic applications. Inspired by the aforementioned exciting success [16] and the robust nature of antimony compounds, we employ systematic, first-principles calculations based on density functional theory (DFT) to study the stability and diverse properties of symmetrical $Sb_2X_2O$ (X = S, Se, Te) monolayers and our introduced, highly asymmetric Janus $Sb_2SSeO$ monolayer. First, we establish the structural integrity of the systems through a combination of formation energy, phonon dispersion, elastic constants, and ab initio molecular dynamics (AIMD) simulations. This confirms that, except for $Sb_2Te_2O$, the other systems show excellent thermodynamic and dynamical stability for prospective experimental realization. We confirm the reliability of our hybrid functional calculation by reproducing the previously reported experimental band gap. Our electronic findings confirm that the band gaps in $Sb_2Se_2O$ and $Sb_2SSeO$ are indirect. Furthermore, we confirm that carrier transport in these 2D systems is characterized by significant anisotropy and substantial tunability under biaxial strain. Finally, we demonstrate that the calculated band alignment is ideal for overall water splitting under neutral conditions. Our extensive first-principles analysis highlights that these novel monolayers are high-performance candidates for next-generation photocatalytic and optoelectronic nanodevices.

**METHODS**

First-principles calculations based on density functional theory (DFT) were carried out within periodic boundary conditions [17] using the Vienna Ab initio Simulation Package (VASP) [18]. The exchange-correlation effects were treated within the generalized gradient approximation (GGA) of Perdew-Burke-Ernzerhof (PBE) [19], while long-range dispersion interactions were included through the semi-empirical Grimme DFT-D3 scheme [20]. The valence electron states were expanded in a plane-wave basis set with a kinetic energy cutoff of 500 eV, and the interaction between valence and core electrons was described by the projector augmented-wave (PAW) method [21]. For electronic structure calculations, the total energy was converged to $10^{-6}$ eV, and structural optimizations were performed until the residual atomic forces were less than 0.02 eV/Å. A vacuum layer of 20 Å was added perpendicular to the monolayer's plane to prevent interactions between periodic images. Dipole corrections were applied in all slab calculations to remove artificial electrostatic interactions between periodic images along the out-of-plane direction. The electronic structures were calculated using the Heyd-Scuseria-Ernzerhof (HSE06) [22] hybrid functional, since the PBE functional underestimates the band gap. Spin-orbit coupling (SOC) was also included on top of the PBE and HSE06 functionals to examine its impact on the band structure and band edge positions, as SOC plays a crucial role in materials containing heavy elements, such as Sb. The optical properties were calculated from the frequency-dependent dielectric matrix, $\varepsilon_{\alpha\beta}(\omega)$, using the HSE06 functional, based on a sum-over-states approach. The number of empty conduction bands was converged for each structure to ensure accurate results. The absorption coefficient was then obtained from the dielectric matrix via the Kramers–Kronig transformation [23]:

$$\alpha_{\alpha\beta}(\omega) = \frac{2\omega k_{\alpha\beta}(\omega)}{c} = \frac{\omega Im(\varepsilon_{\alpha\beta}(\omega))}{c n_{\alpha\beta}(\omega)} \qquad (1)$$

Here, c is the speed of light in a vacuum. At the same time, $n_{\alpha\beta}$ and $k_{\alpha\beta}$ are the real and imaginary parts of the complex refractive index, corresponding to the refractive index and extinction coefficient, respectively. Brillouin zone (BZ) integration was performed using a 3×4×1 k-point mesh for structural optimization, a denser 12×16×1 mesh for electronic structure calculations, and an even denser 18×24×1 mesh for optical property calculations. The charge transfer was evaluated using Bader charge analysis [24] at the same level of theory.

Ab initio molecular dynamics (AIMD) simulations were performed in the canonical (NVT) ensemble using a Langevin thermostat at 500 K, with a time step of 1.0 fs, on a 2×2 supercell and a 2×2×1 k-point mesh. Phonon dispersion relations were obtained via the moment tensor potential (MTP) approach [25,26]. The training dataset was generated from AIMD simulations performed over 750 time steps with temperatures gradually increased from 1 to 1000 K. For each structure, MTPs with a cutoff radius of 5.5 Å were trained using half of the AIMD data, following our previous protocol. The trained MTPs were then employed to compute phonon dispersion relations using the small displacement method on 3×3×1 supercells as implemented in the PHONOPY package [27], consistent with our earlier work [26]. The elastic properties of the monolayers were obtained using the optimized high-efficiency strain-matrix sets (OHESS) method, as implemented in the ElasTool package [28].

We used DFPT, as implemented in VASP, to calculate the permittivity tensor of the monolayer unit cell, from which the in-plane ($\varepsilon\|$) and out-of-plane ($\varepsilon\perp$) dielectric constants were extracted. Since VASP employs a plane-wave basis, the unit cells are periodic in all three directions. For 2D monolayers, the supercell dielectric values include a vacuum contribution that must be removed to obtain the intrinsic dielectric constants of the layer. We rescaled the supercell dielectric constants following the same procedure as for bulk materials. Following Ref. [29], the dielectric constants were calculated using the following equations:

$$\varepsilon_{2D,\perp} = \left[1 + \frac{h}{t}\left(\frac{1}{\varepsilon_{SC,\perp}} - 1\right)\right]^{-1} \quad (2)$$

$$\varepsilon_{2D,\|} = 1 + \frac{h}{t}\left(\varepsilon_{SC,\|} - 1\right) \quad (3)$$

Here, $h$ is the height of the supercell, and $t$ is the thickness of the monolayer, which is determined from the average interlayer distance of the bilayer as shown in Figure S1. Because this method is sensitive to vacuum size, a 30 Å vacuum layer was used for the dielectric constant calculations.

The carrier effective mass (m*) was calculated using two approaches. In the first method, m* was obtained from the curvature of the band structure near the band edges through a parabolic fitting by $m^* = \hbar^2 \left[\frac{\partial^2 E(k)}{\partial k^2}\right]^{-1}$, where $\hbar$ is the reduced Planck constant, $k$ is the wave vector, and $E(k)$ is the band edge energy. However, this parabolic model has two main limitations: (i) the proximity of other band extrema in energy and (ii) the challenge of choosing appropriate crystallographic directions, which restricts its accuracy [30]. In the second method, the effective mass was derived from the carrier mobility using $m^* = \frac{e\tau}{\mu}$, where $\tau$ is the relaxation time (here $\tau$ = 10 fs) and $\mu$ is the carrier mobility, thereby overcoming the above limitations. The carrier mobility was calculated using the AMSET code [31], which requires the band structure, wavefunction coefficients, elastic constants, deformation potentials, dielectric constants, and polar-phonon frequencies as input. The AMSET calculations were performed on top of the HSE06 band gap using an 18×24×1 mesh. To determine the feasibility of these 2D semiconductors for scalable photocatalytic water splitting applications, we calculated and analyzed the solar-to-hydrogen (STH) conversion efficiency ($\eta_{STH}$), along with the energy conversion efficiencies of light absorption ($\eta_{abs}$), carrier utilization ($\eta_{cu}$), and corrected STH ($\eta'_{STH}$), as described in the Supporting Information.

## RESULTS AND DISCUSSION

**Stability, structural, and elastic properties**

The monolayer structures were derived from the bulk $Sb_2S_2O$ in the triclinic phase. The optimized bulk structure with its lattice parameters is shown in Figure S2. The optimized lattice parameters of the bulk $Sb_2S_2O$ obtained in our calculations (a = 5.89 Å, b = 8.19 Å, c = 10.74 Å; α = 100.83°, β = 102.83°, γ = 110.59°) show excellent agreement with the experimental values (a = 5.78 Å, b = 8.14 Å, c = 10.70 Å; α = 101.02°, β = 102.76°, γ = 110.66°) [32]. In contrast, our results differ from previously reported DFT values (a = 6.06 Å, b = 8.40 Å, c = 11.07 Å) [16], which may be attributed to the relatively coarse k-point mesh (3×2×1) used in that study for bulk calculations, whereas we employed a denser 6×5×3 mesh. Since the bulk $Sb_2S_2O$ is a van der Waals structure, a single layer was extracted and optimized to obtain its lattice parameters. The S atoms were then replaced with Se and Te to construct the $Sb_2Se_2O$ and $Sb_2Te_2O$ monolayer structures. Due to the presence of imaginary frequencies in the phonon dispersion of the $Sb_2Te_2O$ monolayer (Figure S3), indicating its dynamical instability, we restrict our study to the structural and optoelectronic properties of the $Sb_2S_2O$ and $Sb_2Se_2O$ monolayers and their Janus counterpart. Figure 1 (a, b) shows the top and side views of the atomic structures of $Sb_2X_2O$ (X = S, Se) monolayers, along with their lattice parameters. The structure of the Janus $Sb_2SSeO$ monolayer is depicted in Figure S4. The primitive cell of $Sb_2X_2O$ monolayers is a triclinic lattice formed by 20 atoms (8 Sb, 8 X, and 4 O atoms) with space group number 2 ($P\bar{1}$). In the crystal structure, there are two distinct types of Sb atoms: one type is coordinated by five X atoms, while the other is bonded to three O atoms and a single X atom. The calculated lattice parameters *a (b)* of the $Sb_2S_2O$, $Sb_2Se_2O$, and $Sb_2SSeO$ monolayers are 11.12 (8.28) Å, 11.44 (8.36) Å, and 11.37 (8.34) Å, respectively, as listed in Table 1. The difference between the maximum and minimum z-coordinates of the X atoms in the $Sb_2S_2O$ monolayer is 3.22 Å, while that of the $Sb_2Se_2O$ monolayer is 0.35 Å larger. Additionally, all Sb-X and Sb-O bond lengths in $Sb_2Se_2O$ are longer than those in $Sb_2S_2O$, with the increase being particularly pronounced for the Sb-X bond. Bader charge analysis shows that in the $Sb_2S_2O$ monolayer, Sb atoms coordinated by five S atoms lose an average of 1.16 e, while Sb atoms bonded to three O atoms and one S atom lose 1.6 e. Consequently, the S atoms gain 0.8 e on average, and the O atoms gain 1.16 e. In the $Sb_2Se_2O$ monolayer, Sb atoms coordinated by five Se atoms lose 0.93 e on average, whereas Sb atoms bonded to three O atoms and one Se atom lose 1.52 e, resulting in average gains of 0.64 e for Se and 1.17 e for O. These results further indicate that the Sb-Se bond has a stronger covalent character compared to the Sb-S bond. It is worth noting that the observed charge redistribution arises from the differences in electronegativity: 2.04 for Sb, 2.58 for S, 2.55 for Se, and 3.44 for O. These results indicate that the Sb-X bonds are less ionic than the Sb-O bonds, consistent with the lower electronegativity of X atoms compared to O, as further illustrated by the electron localization function (ELF) in Figure 1(c, d, e).

We next investigate the energy required to isolate $Sb_2X_2O$ (X = S, Se) monolayers from their multilayered crystals via mechanical exfoliation. The exfoliation process was modeled by first optimizing the six-layer structures and then gradually separating the topmost layer toward the vacuum in 0.1 Å increments while monitoring the corresponding change in total energy (see Figure 1(f) inset). As shown in Figure 1(f), the exfoliation energy initially rises steeply at small separations, then progressively approaches a plateau, ultimately converging near ~8 Å for both monolayers. The resulting cleavage energies, 0.36 J m$^{-2}$ for $Sb_2S_2O$ and 0.40 J m$^{-2}$ for $Sb_2Se_2O$, confirm the weak interlayer coupling in these materials. These values are comparable to the experimentally measured cleavage energy of graphene (0.37 J m$^{-2}$) [33] and are substantially lower than those of more strongly bound nanosheets like molybdenum disulfide (MoS$_2$, ≈0.6 J m$^{-2}$) [34]. This comparison places $Sb_2X_2O$ firmly within the class of materials that can be readily isolated using standard mechanical exfoliation. The slight increase in exfoliation energy from S to Se follows the expected trend of enhanced van der Waals interactions with increasing chalcogen atomic weight. Overall, the relatively low cleavage energies highlight the strong feasibility of experimentally isolating $Sb_2X_2O$ monolayers and reinforce their promise within the broader family of exfoliable 2D materials.

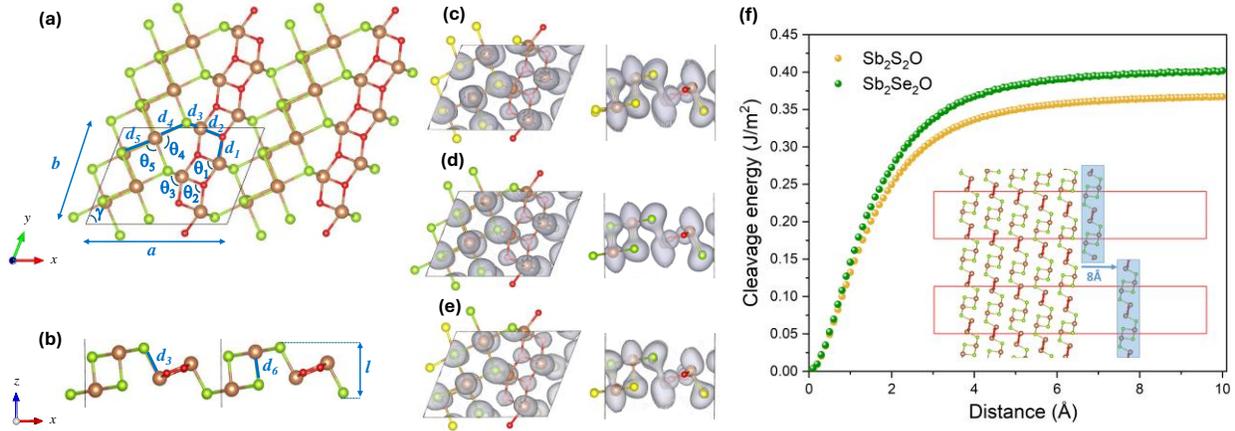

Figure 1. Top (a) and side (b) views of the $Sb_2X_2O$ (X = S, Se) monolayer atomic structure, with the unit cell outlined by a black solid line and lattice parameters indicated. Panels (c), (d), and (e) show the electron localization function (ELF) isosurfaces for $Sb_2S_2O$, $Sb_2Se_2O$, and $Sb_2SSeO$, respectively. (f) Cleavage energy profiles of $Sb_2S_2O$ and $Sb_2Se_2O$ monolayers extracted from their six-layer bulk structures as a function of interlayer separation. Atom colors: orange = Sb, green = Se, yellow = S, and red = O.

Table 1. Structural parameters of $Sb_2X_2O$ (X = S, Se) monolayers, as illustrated in Figure 1. Reported values include the lattice constants (a, b), lattice angle (γ), Sb-X and Sb-O bond lengths ($d_1$-$d_6$), bond angles ($\theta_1$-$\theta_5$), and cohesive energies. The difference between the maximum and minimum z-coordinates of the X atoms (l). Lattice parameters and bond lengths are given in Angstroms (Å), bond angles are expressed in degrees (°), and cohesive energies in eV. For the Janus $Sb_2SSeO$ monolayer, the values outside and inside the parentheses correspond to the Sb-S and Sb-Se parameters, respectively.

| System | | | $Sb_2S_2O$ ML | $Sb_2Se_2O$ ML | $Sb_2SSeO$ ML |
|---|---|---|---|---|---|
| **Lattice parameters** | | a | 11.12 | 11.44 | 11.37 |
| | | b | 8.28 | 8.36 | 8.34 |
| | | γ | 68.17 | 68.62 | 68.49 |
| | | l | 3.22 | 3.57 | 3.51 |
| **Bond lengths** | | $d_1$ | 2.16 | 2.19 | 2.20 |
| | | $d_2$ | 2.04 | 2.05 | 2.05 |
| | | $d_3$ | 2.50 | 2.64 | 2.50 (2.66) |
| | | $d_4$ | 2.58 | 2.87 | 2.57 (2.88) |
| | | $d_5$ | 2.63 | 2.98 | 2.65 (2.98) |
| | | $d_6$ | 2.46 | 2.58 | 2.45 (2.57) |
| **Bond angles** | | $\theta_1$ | 72.56 | 72.79 | 72.68 |
| | | $\theta_2$ | 106.84 | 107.20 | 107.86 |
| | | $\theta_3$ | 87.14 | 88.15 | 87.11 (91.02) |
| | | $\theta_4$ | 100.24 | 93.35 | 100.31 (92.23) |
| | | $\theta_5$ | 86.94 | 88.85 | 94.52 (89.86) |
| **Formation energy** | | | -11.00 | -10.64 | -10.80 |

To assess the stability of these novel monolayers, we analyzed the formation energy, phonon dispersion, and AIMD simulations at 300 K and 500 K. The calculated formation energies of $Sb_2S_2O$, $Sb_2Se_2O$, and $Sb_2SSeO$ monolayers are -11.00, -10.64, and -10.80 eV/atom, respectively (see Supporting Information S7), confirming their energetic stability. These values are lower (more negative) than those reported for $Sb_2X_3$ (X = S, Se, Te) single layers [35], indicating that $Sb_2X_2O$ monolayers are comparatively more stable. Furthermore, in both $Sb_2X_2O$ and $Sb_2X_3$ 2D systems, the stability increases as the chalcogen atom becomes lighter. Phonon dispersion calculations across the entire Brillouin zone were carried out to further evaluate the dynamical stability of these novel 2D materials. As shown in Figure 2 (a, b, c), the absence of imaginary frequencies in the phonon spectra confirms their stability. In the $Sb_2S_2O$ monolayer, the optical phonon frequencies are higher than in $Sb_2Se_2O$, mainly because sulfur is lighter than selenium and the Sb-S bonds are stronger, which together reduce the effective reduced mass and increase the bond stiffness. In contrast, the heavier Se atoms and weaker Sb-Se bonds in the $Sb_2Se_2O$ monolayer lead to lower phonon frequencies. In addition, the highest phonon frequencies of monolayer $Sb_2S_2O$, $Sb_2Se_2O$, and $Sb_2SSeO$ reach 18.74 THz (625 cm$^{-1}$). 18.22 THz (608 cm$^{-1}$), and 18.50 THz (617 cm$^{-1}$) along the $H_1$-H direction, respectively. These values are higher than those of silicene (580 cm$^{-1}$) [36] and black phosphorene (540 cm$^{-1}$) [37], providing further evidence for the stability of these novel 2D systems. Figure 2(d, e, f) shows the total energy fluctuations from the AIMD simulations of the studied monolayers at 300 K and 500 K, along with top and side view snapshots of the structures after 10 ps. Analysis of the AIMD trajectories indicates that the structures remain intact at both 300 and 500 K, with stable energy and temperature profiles, confirming the thermal stability of the $Sb_2X_2O$ monolayers. Based on the combined results of formation energy, phonon dispersion, and AIMD simulations, we conclude that these nanostructures are stable and could potentially be synthesized experimentally.

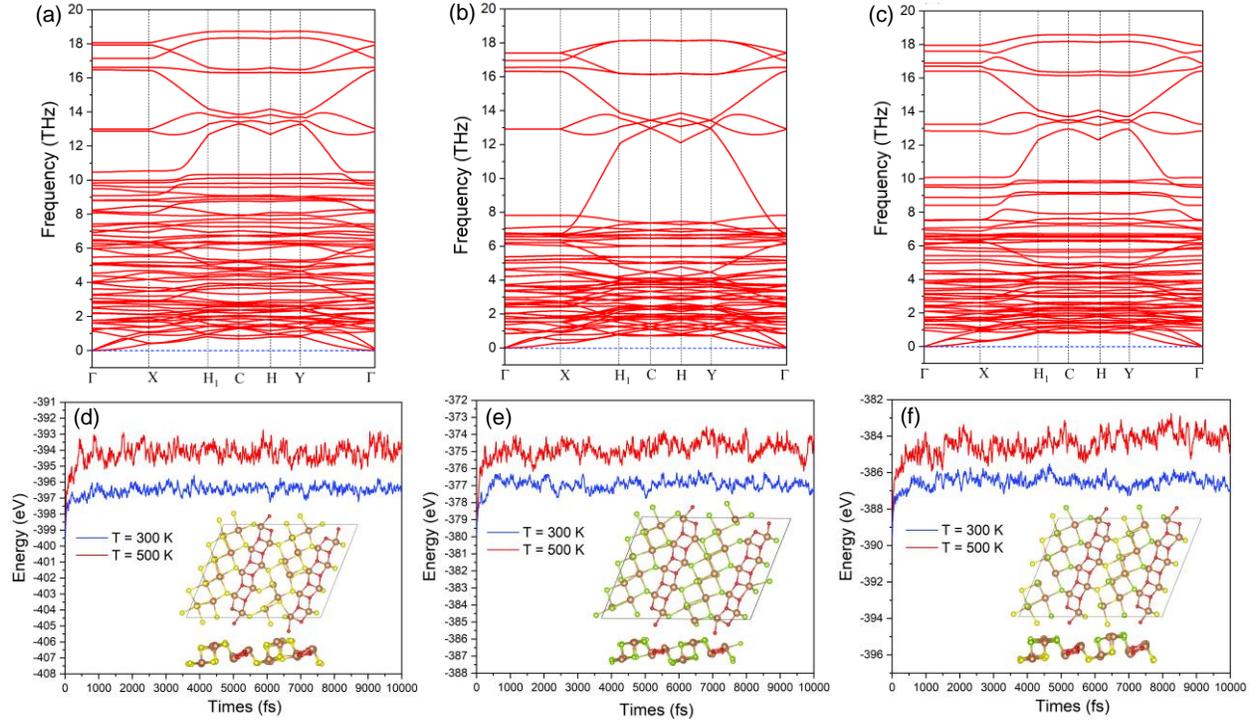

Figure 2. Phonon dispersion of single-layer $Sb_2S_2O$ (a), $Sb_2Se_2O$ (b), and $Sb_2SSeO$ (c). Ab initio molecular dynamics (AIMD) simulations for $Sb_2S_2O$ (d), $Sb_2Se_2O$ (e), and $Sb_2SSeO$ (f) monolayers at 300 and 500 K. The total energy profiles over the simulation time indicate thermal stability, with negligible structural distortion observed. The top and side views correspond to the optimized configurations after 10 ps of simulation at 500 K. Atom colors: orange = Sb, yellow = S, green = Se, red = O.

Next, we examine the elastic properties of these 2D materials. The angular dependence of Young's modulus, Poisson's ratio, shear modulus, and stiffness constants for both monolayers is illustrated in Figure 3, while the corresponding elastic parameters along the x- and y-directions are summarized in Table 2. Our results reveal that the elastic properties of all studied monolayers are strongly anisotropic, which originates from their inherently anisotropic crystal structure. The Young's modulus of monolayer $Sb_2S_2O$ varies from 35.89 N/m to 25.00 N/m as the angle θ changes, while that of $Sb_2Se_2O$ ranges from 40.09 N/m to 25.45 N/m. For the Janus $Sb_2SSeO$ monolayer, the corresponding values lie between 37.67 N/m and 23.70 N/m. These values are significantly smaller than those of most reported 2D materials, such as graphene (342 N/m) [38], and $MoS_2$ (127 N/m) [39]. The comparatively low Young's modulus suggests that $Sb_2X_2O$ monolayers are much softer and more flexible, which could be advantageous for applications requiring mechanical deformability, such as flexible and wearable electronic or optoelectronic devices.

Poisson's ratio is defined as the ratio of the transverse strain to the longitudinal strain under uniaxial stress. As shown in the Figure. 3(b), the monolayer $Sb_2S_2O$ exhibits a maximum Poisson's ratio of 0.07 along the x-direction (θ = 90°) and 0.05 along the y-direction (θ = 0°). In contrast, the $Sb_2Se_2O$ monolayer shows considerably larger values, reaching 0.26 along the x-direction (θ = 90°) and 0.16 along the y-direction (θ = 0°). For the Janus $Sb_2SSeO$ monolayer, the corresponding values lie between the two, with 0.15 along the x-direction and 0.09 along the y-direction. The relatively high Poisson's ratio of $Sb_2Se_2O$ suggests that it experiences a stronger lateral contraction or expansion response to applied strain compared to $Sb_2S_2O$ and $Sb_2SSeO$. Importantly, all values remain positive, confirming that these novel 2D systems are not auxetic materials. Moreover, the strong anisotropy in Poisson's ratio of the $Sb_2Se_2O$ monolayer shows that the heavier Se atoms make the lattice more sensitive to in-plane deformations. Our results show that $Sb_2S_2O$ exhibits relatively low and weakly anisotropic Poisson's ratio values, while $Sb_2Se_2O$ displays pronounced anisotropy with significantly higher values. For comparison, graphene has an almost isotropic Poisson's ratio of ~0.16 [40], and $MoS_2$ has a value of 0.26 [41]. These findings further confirm the structural flexibility and direction-dependent mechanical properties of these novel monolayers. From Figure 3(c), the maximum shear modulus of $Sb_2S_2O$, $Sb_2Se_2O$, and $Sb_2SSeO$ monolayers is 16.73 N/m, 15.93 N/m, and 16.42 N/m, occurring at θ = 70°, while the minimum values are 11.90 N/m, 10.93 N/m, and 10.85 N/m at θ = 30°. The higher shear modulus of $Sb_2S_2O$ indicates that it is more resistant to in-plane shear deformation compared to $Sb_2Se_2O$. This difference reflects the anisotropic bonding environment in the monolayers. While $Sb_2S_2O$ has slightly lower Young's modulus along certain directions, its higher shear modulus at θ = 70° suggests that the Sb-S bonding network provides stronger resistance to shear deformation in specific directions. The maximum stiffness constants of the $Sb_2S_2O$, $Sb_2Se_2O$, and $Sb_2SSeO$ monolayers are approximately 35.00 N/m, 39.00 N/m, and 36.50 N/m, respectively, occurring at θ = 90°. These values indicate that all monolayers are mechanically stiffer along this direction, reflecting the anisotropic nature of their bonding networks. The slightly higher stiffness in the $Sb_2Se_2O$ monolayer suggests stronger resistance to deformation along θ = 90°, which can be attributed to the directional arrangement of Sb-Se and Sb-O bonds that govern in-plane rigidity.

The mechanical stability of the monolayer was evaluated using the Born-Huang criteria [42]. For such systems, mechanical stability requires the elastic stiffness constants to satisfy the following conditions: (i) $C_{11} > 0$, (ii) $C_{22} > 0$, (iii) $C_{66} > 0$, and (iv) $C_{11}C_{22} - C_{12}^2 > 0$. These inequalities ensure the material's resistance to both longitudinal and shear deformations. Since the $Sb_2X_2O$ monolayers fulfil all of these criteria, they can be considered mechanically stable.

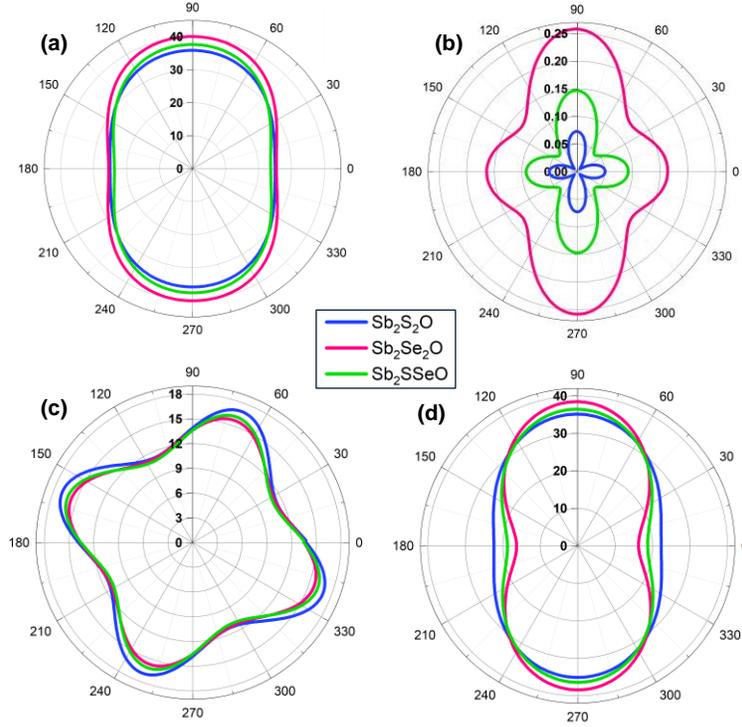

Figure 3. Angular dependence of the elastic properties for $Sb_2S_2O$, $Sb_2Se_2O$, and $Sb_2SSeO$ monolayers: (a) Young's modulus, (b) Poisson's ratio, (c) shear modulus, and (d) stiffness constant.

Table 2. Elastic parameters of $Sb_2S_2O$, $Sb_2Se_2O$, and $Sb_2SSeO$ monolayers.

| System | | | $Sb_2S_2O$ ML | $Sb_2Se_2O$ ML | $Sb_2SSeO$ ML |
|---|---|---|---|---|---|
| Elastic tensor (N/m) | | $C_{11}$ | 25.09 | 26.57 | 24.02 |
| | | $C_{12}$ | 1.82 | 6.86 | 3.53 |
| | | $C_{22}$ | 36.02 | 41.85 | 38.18 |
| | | $C_{66}$ | 15.44 | 14.91 | 15.03 |
| Young's modulus (N/m) | | xx | 35.89 | 40.09 | 37.67 |
| | | yy | 25.00 | 25.45 | 23.70 |
| Shear modulus (N/m) | | xx | 16.73 | 15.93 | 16.42 |
| | | yy | 11.90 | 10.93 | 10.85 |
| Poisson's ratio | | xx | 0.07 | 0.26 | 0.15 |
| | | yy | 0.05 | 0.16 | 0.09 |
| Stiffness constant (N/m) | | xx | 19.35 | 27.01 | 22.08 |
| | | yy | 13.16 | 15.22 | 13.06 |

**Electronic structure and prerequisites for photocatalytic water splitting**

To investigate the electronic properties of these novel 2D materials, we calculated the band structures, total density of states (DOS), and projected DOS. Figure 4 presents the band structures along high-symmetry directions, together with the total and partial DOS of $Sb_2S_2O$, $Sb_2Se_2O$, and $Sb_2SSeO$ monolayers, obtained using the HSE06 functional both with and without SOC. Our results reveal that the $Sb_2S_2O$ monolayer is a direct bandgap semiconductor with an energy gap of 2.80 eV (2.74 eV) as calculated using the HSE06 (HSE06+SOC) functional. Both the valence band maximum (VBM) and the conduction band minimum (CBM) are located at the X point. The band gap of the $Sb_2S_2O$ monolayer is larger than its experimental bulk band gap of 2.04 eV [16], highlighting the quantum confinement effect in the reduced-dimensional structure. It is worth noting that the calculated band gap of bulk $Sb_2S_2O$ in our work is 2.00 eV using the HSE06 functional, which is in excellent agreement with the experimental value of 2.04 eV [16]. The corresponding band structure and DOS are presented in Figure S5. This agreement confirms the high accuracy and reliability of our computational approach. In contrast, the $Sb_2Se_2O$ and $Sb_2SSeO$ monolayers exhibit indirect band gaps of 2.24 eV (2.15 eV with SOC) and 2.44 eV (2.35 eV with SOC), respectively. In both cases, the valence band maximum (VBM) is located along the X-$H_1$ direction, while the conduction band minimum (CBM) remains at the X point. The partial orbital DOS (Figure S6), together with the partial charge densities of the VBM and CBM (Figure 4a-c), reveals that the valence band maximum is predominantly composed of the *p* orbitals of X and O atoms, while the conduction band minimum originates mainly from Sb 5*p* orbitals with minor contributions from X *p* states. This orbital distribution confirms the charge-transfer insulator nature of $Sb_2X_2O$ monolayers. As a result, SOC exerts only a negligible influence on their electronic structures, since the band edges are largely determined by the lighter-element *p* states (X and O), which are far less sensitive to spin-orbit coupling than the heavier Sb orbitals. Our results confirm that these novel 2D systems possess suitable electronic band gaps for efficient visible-light absorption, as an ideal photocatalyst typically requires a band gap in the range of 2.1-2.8 eV, depending on the targeted application (e.g., water splitting or $CO_2$ reduction) [30].

The work function is calculated as $\emptyset = E_{vaccum} - E_F$, where $E_{vacuum}$ is the vacuum level, obtained from the planar-averaged electrostatic potential shown in Figure 4(d-f), and $E_F$ is the Fermi level. The calculated work functions at the PBE level are 4.47, 4.30, and 4.35 eV for $Sb_2S_2O$, $Sb_2Se_2O$, and $Sb_2SSeO$ monolayers, respectively, while the corresponding values obtained using the HSE06 functional are 4.94, 4.70, and 4.82 eV. These results reveal a clear trend: the work function decreases as the electronegativity of the chalcogen atom decreases. Table 3 summarizes the calculated electronic band gaps and work functions of all studied monolayers, obtained using PBE and HSE06 functionals, both with and without SOC.

For an efficient overall water-splitting photocatalyst, the CBM must lie above the hydrogen reduction potential (−4.44 eV at pH = 0), while the VBM should lie below the oxygen oxidation potential (−5.67 eV at pH = 0). As shown in Figure 4(d-f), all studied monolayers satisfy these criteria, with their CBM and VBM straddling the redox potentials of $H^+/H_2$ and $H_2O/O_2$, confirming their ability to drive both HER and OER. Furthermore, since the redox potentials shift with pH according to the Nernst equation (59 mV per pH unit) [43], we also examined the alignment at pH = 7. The results show that the band edges of $Sb_2S_2O$, $Sb_2Se_2O$, and $Sb_2SSeO$ remain well-positioned relative to the water redox potentials, demonstrating their robustness as photocatalysts under neutral conditions. According to previous literature [15], the potential of photogenerated electrons for the hydrogen reduction reaction ($U_e$) is defined as the energy difference between the CBM and the hydrogen reduction potential, while the potential of photogenerated holes for water oxidation ($U_h$) is calculated as the energy difference between the VBM and the hydrogen reduction potential. The reduction potential of $H^+/H_2$ varies with pH according to the Nernst equation:

$$E^{red}_{H^+/H_2} = -4.44 + pH \times 0.059 \ (eV) \qquad (4)$$

Moreover, in the case of the Janus monolayer, the reduction potential is further enhanced by the electrostatic potential difference (Δφ), which is calculated to be 0.06 eV. This behavior is consistent with Yang's theory [44], and can be expressed as:

$$E^{red\ (Janus)}_{H^+/H_2} = E^{red}_{H^+/H_2} + \Delta\varphi \qquad (5)$$

In a neutral environment (pH = 7), the calculated $U_e$ values for $Sb_2S_2O$, $Sb_2Se_2O$, and $Sb_2SSeO$ monolayers are 0.88, 0.70, and 0.81 V, respectively, while the corresponding $U_h$ values are 1.92, 1.55, and 1.75 V. The fact that both $U_e$ and $U_h$ are positive for all cases confirms that these monolayers possess sufficient driving forces for both hydrogen evolution and oxygen evolution reactions, highlighting their potential as efficient photocatalysts for overall water splitting under neutral conditions. For real-world photocatalytic applications, it is essential to evaluate the stability of candidate materials in aqueous solutions under illumination. Using the methodology proposed by Chen et al. [45], we determined the thermodynamic oxidation potential ($\Phi^{ox}$) and reduction potential ($\Phi^{re}$) of the studied monolayers. The corresponding methods are summarized in the Supporting Information (Section S7). As shown in Figure 4(d-f), the calculated $\Phi^{ox}$ values (red lines) for all studied monolayers lie below the oxidation potential of $O_2/H_2O$, while their $\Phi^{re}$ values (blue lines) are above the reduction potential of $H^+/H_2$. This alignment indicates that the photogenerated charge carriers are more likely to participate in water-splitting reactions rather than oxidizing or reducing the photocatalyst itself [45]. Therefore, these novel monolayers are expected to exhibit strong resistance against photoinduced corrosion.

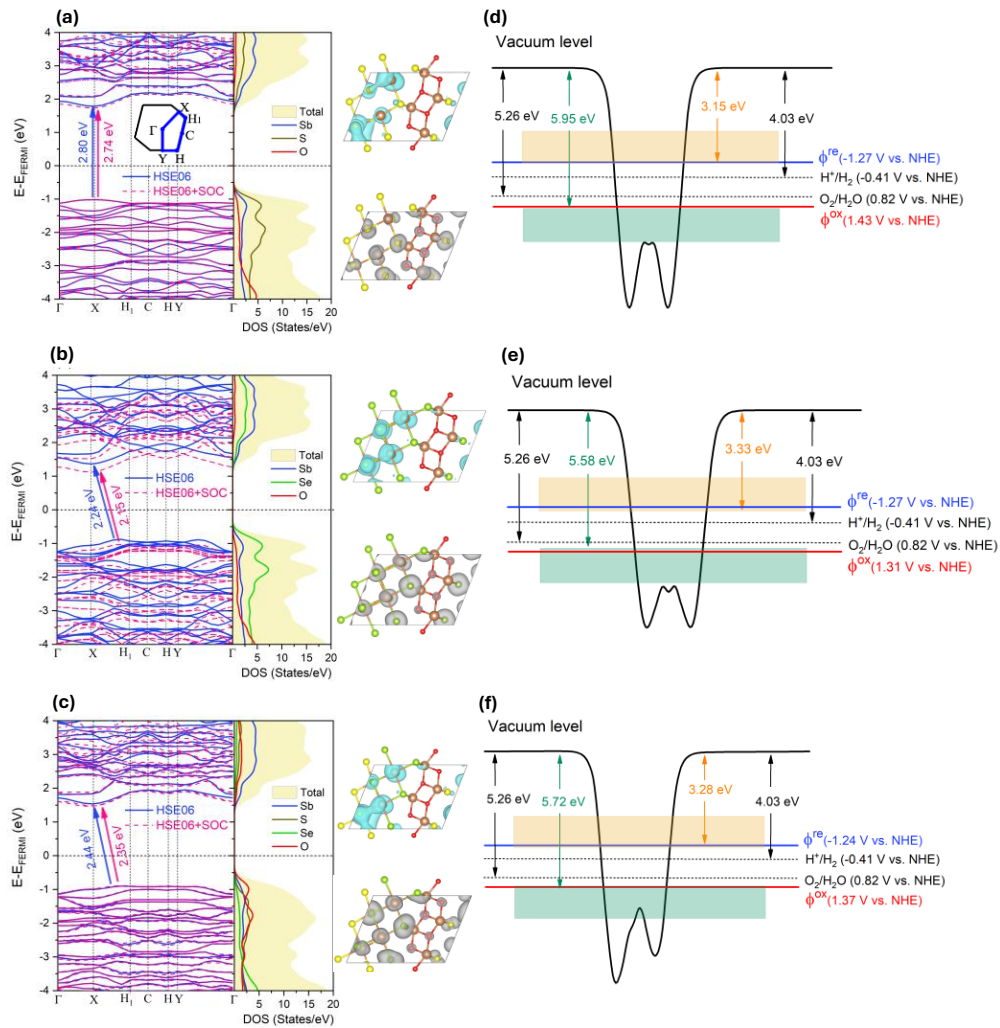

Figure 4. Electronic band structure, total density of states (DOS), and partial DOS for $Sb_2S_2O$ (a), $Sb_2Se_2O$ (b), and $Sb_2SSeO$ (c) monolayers calculated using the HSE06 functional. The band structures, including spin-orbit coupling (SOC) at the HSE06 level, are shown as dashed pink lines. The Fermi energy is set to zero. Partial charge densities of the VBM (gray isosurfaces) and CBM (cyan isosurfaces) are also shown for each monolayer. A schematic of the Brillouin zone, along with the high-symmetry k-points used for the band structure calculations, is provided in (a). Panels (d), (e), and (f) present the planar-averaged electrostatic potential and band edge

positions of $Sb_2S_2O$, $Sb_2Se_2O$, and $Sb_2SSeO$ monolayers, respectively, relative to the water redox potentials at pH = 7 (black dots), based on HSE06 calculations.

Table 3. Calculated electronic band gaps and work functions for $Sb_2S_2O$, $Sb_2Se_2O$, and $Sb_2SSeO$ monolayers using different exchange-correlation functionals.

| System | Band gap (eV) | | | | Work function (eV) | | | |
|---|---|---|---|---|---|---|---|---|
| | PBE | PBE+SOC | HSE06 | HSE06+SOC | PBE | PBE+SOC | HSE06 | HSE06+SOC |
| $Sb_2S_2O$ | 1.95 | 1.84 | 2.80 | 2.74 | 4.47 | 4.44 | 4.94 | 4.94 |
| $Sb_2Se_2O$ | 1.55 | 1.41 | 2.24 | 2.15 | 4.30 | 4.35 | 4.70 | 4.54 |
| $Sb_2SSeO$ | 1.67 | 1.58 | 2.44 | 2.35 | 4.35 | 4.39 | 4.82 | 4.82 |

**Dielectric constant and absorption coefficient**

Another key parameter influencing the photocatalytic activity of semiconductors is the dielectric constant, which generally needs to exceed a value of 10 to ensure efficient photocatalytic performance [30]. The relative dielectric constant ($\varepsilon_r$) is composed of two contributions: the electronic part ($\varepsilon_\infty$), which reflects the response of the electron density, and the vibrational part ($\varepsilon_{vib}$) from lattice vibrations, such that $\varepsilon_r = \varepsilon_\infty + \varepsilon_{vib}$. Figure S7(a, b) together with Table S2 present the calculated electronic ($\varepsilon_\infty$) and relative ($\varepsilon_r$) dielectric constants for all the studied monolayers. In all studied monolayers, $\varepsilon_\infty$ and $\varepsilon_{vib}$ exhibit pronounced anisotropy, with smaller values along the out-of-plane axis relative to the in-plane axes. Moreover, $\varepsilon_\infty$ of $Sb_2Se_2O$ and $Sb_2SSeO$ monolayers is higher than that of $Sb_2S_2O$, which can be attributed to two related factors: the greater polarizability of Se compared to S, and the stronger covalent character of the Sb-Se bond relative to Sb-S, as previously discussed. The relatively large $\varepsilon_{vib}$ values observed in these 2D systems arise from the higher ionic character of the oxide atoms compared to the sulfide and selenide atoms.

We next investigate the optical absorption properties of the $Sb_2X_2O$ and $Sb_2SSeO$ monolayers under in-plane polarization using the DFT-HSE06 method. The absorption spectra as a function of photon energy are shown in Figure S7c, while Figure S7d presents a comparative view in terms of wavelength within the UV-visible range (300-600 nm). The calculated absorption coefficients reveal pronounced anisotropy in the in-plane optical response. Importantly, all studied monolayers exhibit strong absorptance in the UV-visible region of sunlight. For systems containing Se, the absorptance is further enhanced, consistent with their narrower electronic band gaps. Remarkably, absorptance values reach up to ~18% within the 2.0–4.0 eV (300–500 nm) range, which is significant considering the atomically thin nature of these materials. For comparison, monolayer $MoS_2$ with a thickness of ~3.1 Å shows absorptance between 5-10% at peak excitonic resonances [46], while graphene absorbs only ~2.3% per layer [47]. Such levels in conventional 2D materials typically require strong excitonic effects or additional photonic engineering. In contrast, the $Sb_2X_2O$ and $Sb_2SSeO$ monolayers achieve comparable or higher absorptance without relying on multilayer stacking or resonant structures. The combination of relatively high dielectric constants and strong optical absorptance highlights the promising potential of these novel monolayers for efficient photocatalytic applications.

**Carrier effective masses and mobilities**

As described in the methods section, we employed two approaches, the parabolic band model and a method based on carrier mobility, to evaluate the carrier effective masses. The results along the x- and y-directions are shown in Figure 5(a, b). For the pristine $Sb_2S_2O$ monolayer, the electron and hole effective masses obtained from the parabolic model are 0.68 (1.37) $m_e$ and 3.00 (2.81) $m_e$ along the x (y) direction, respectively. In comparison, the values for the $Sb_2Se_2O$ monolayer are 0.67 (1.23) $m_e$ and 0.87 (0.94) $m_e$, while those for the Janus $Sb_2SSeO$ monolayer are 0.75 (1.10) $m_e$ and 1.60 (0.65) $m_e$, for electrons and holes along the x (y) direction, respectively. The carrier effective masses obtained from the carrier mobility method are generally smaller across all systems compared to those derived from the parabolic model. In the $Sb_2S_2O$ monolayer, the electron effective mass is 0.18 $m_e$ along the x-direction and 0.19 $m_e$ along the y-direction, while the hole effective mass is 0.12 $m_e$ (x) and 0.05 $m_e$ (y). For the $Sb_2Se_2O$ monolayer, electrons are much heavier, with effective masses of 1.71 $m_e$ (x) and 0.56 $m_e$ (y), whereas holes remain relatively light at 0.15 $m_e$ (x) and 0.19 $m_e$ (y). In the case of the Janus $Sb_2SSeO$ monolayer, the electron effective masses are 0.37 $m_e$ along x and 0.71 $m_e$ along y, while the hole effective masses are 0.19 $m_e$ and 0.17 $m_e$ along x and y, respectively.

As shown in Figure 5(c-e), all the studied monolayers exhibit high electron and hole mobilities at 300 K under different doping concentrations. For the $Sb_2S_2O$ monolayer at a doping concentration of $1 \times 10^{15}$ $cm^{-2}$, the electron mobility is 98.06 $cm^2/V·s$ along the x-direction and 94.56 $cm^2/V·s$ along the y-direction, while the hole mobility reaches 147.68 $cm^2/V·s$ (x) and 400 $cm^2/V·s$ (y). In the $Sb_2Se_2O$ monolayer, the corresponding electron mobilities are 10.24 $cm^2/V·s$ (x) and 31.65 $cm^2/V·s$ (y), with hole mobilities of 114.34 $cm^2/V·s$ (x) and 95.33 $cm^2/V·s$ (y). For the Janus $Sb_2SSeO$ monolayer, the electron mobilities are 46.97 $cm^2/V·s$ along x and 24.81 $cm^2/V·s$ along y, while the hole mobilities are 91.17 $cm^2/V·s$ (x) and 104.30 $cm^2/V·s$ (y). These results confirm that both the carrier effective masses and mobilities in these novel 2D materials exhibit pronounced anisotropy. It is worth noting that an efficient photocatalyst should possess carrier effective masses below 0.5 $m_e$ to ensure high carrier mobility (>10 $cm^2/V·s$) [30]. This criterion is satisfied by all three studied monolayers, highlighting their strong potential for photocatalytic applications. As shown in Figure 5(f–h), carrier mobility decreases with increasing temperature because of enhanced phonon scattering at higher temperatures, which hinders the transport of charge carriers and reduces their mobility. To place the carrier transport properties of $Sb_2X_2O$ monolayers in context, we compared our calculated mobilities with well-known 2D semiconductors. The obtained electron and hole mobilities are comparable to those of monolayer $MoS_2$ ($\mu_x^e \approx 72$, $\mu_y^e \approx 200$, $\mu_x^h \approx 60$, $\mu_y^h \approx 152$ cm² V⁻¹ s⁻¹) and $WS_2$ ($\mu_x^e \approx 120$, $\mu_x^h \approx 210$ cm² V⁻¹ s⁻¹) [48], and notably exceed those reported for 2D Sb ($\mu_x^e \approx 45$, $\mu_y^e \approx 15$, $\mu_x^h \approx 34$, $\mu_y^h \approx 16$ cm² V⁻¹ s⁻¹) [49]. This comparison highlights that $Sb_2X_2O$ monolayers possess competitive intrinsic carrier mobilities, further reinforcing their suitability for photocatalytic and optoelectronic applications. It is important to note that carrier mobility directly influences the photocatalytic performance of two-dimensional semiconductors. After photoexcitation, electrons and holes must migrate efficiently to the surface active sites before recombination occurs. Higher carrier mobilities enable faster charge transport, reduce electron–hole recombination, and improve charge separation efficiency. As a result, the likelihood that photogenerated carriers participate in the HER or OER reactions increases. Therefore, 2D materials with lower effective masses and higher mobilities generally exhibit better photocatalytic activity. The carrier mobilities obtained for $Sb_2X_2O$ and $Sb_2SSeO$ monolayers, which are comparable to those of well-studied 2D semiconductors such as $MoS_2$ and $WS_2$, indicate that these systems possess suitable charge-transport properties for water-splitting applications.

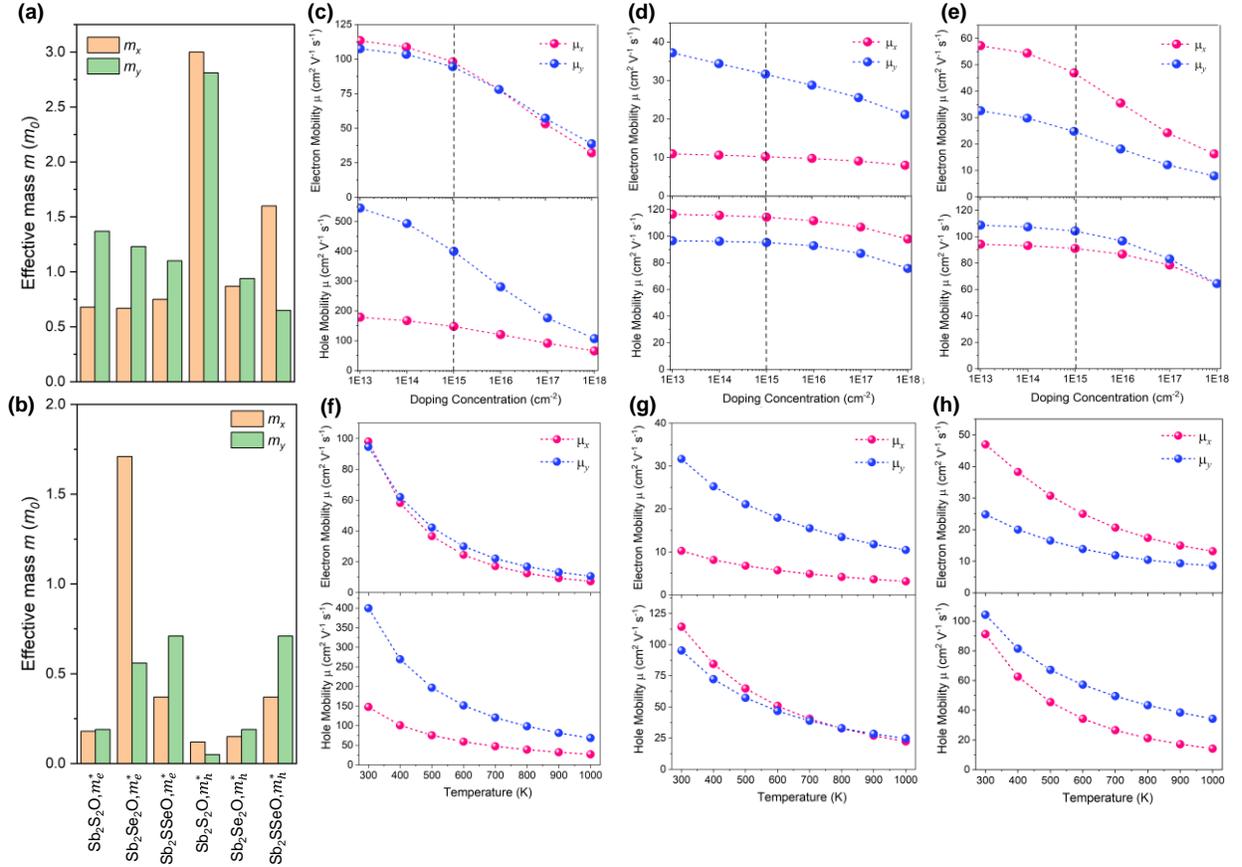

Figure 5. Carrier effective masses for $Sb_2X_2O$ monolayers obtained from $m^* = \hbar^2 \left[\frac{\partial^2 E(k)}{\partial k^2}\right]^{-1}$ (a) and $m^* = \frac{e\tau}{\mu}$ (b). Electron and hole mobilities for $Sb_2S_2O$ (c), $Sb_2Se_2O$ (d), and $Sb_2SSeO$ (e) monolayers at 300 K as a function of doping concentration. Panels (f), (g), and (h) show the temperature dependence of electron and hole mobilities at a fixed doping concentration of $1 \times 10^{15}$ cm$^{-2}$ for $Sb_2S_2O$, $Sb_2Se_2O$, and $Sb_2SSeO$, respectively.

**Effect of biaxial strain on band edge alignment and energy conversion efficiency**

Strain engineering is a powerful strategy for tuning the electronic properties of 2D materials. To investigate its impact, we applied biaxial strain ranging from −6% to 6% to all studied monolayers and examined the resulting evolution of the band gap, CBM, and VBM positions in vacuum, as well as their corresponding energy conversion efficiency at pH = 0 and pH = 7. Figure 6 presents the evolution of the band edge positions and band gaps of $Sb_2S_2O$, $Sb_2Se_2O$, and $Sb_2SSeO$ monolayers under biaxial strain, calculated using the HSE06 functional. The results reveal that the band gap of $Sb_2S_2O$ can be tuned within the range of 2.24-3.07 eV, while $Sb_2Se_2O$ exhibits a tunability from 1.70 to 2.52 eV, and the Janus $Sb_2SSeO$ varies between 1.94 and 2.67 eV. In all the studied monolayers, the CBM remains nearly unchanged under tensile strain, which confirms that the $U_e$ is almost unaffected. In contrast, the VBM shifts to lower (more negative) potentials under tensile strain, leading to an increase in the $U_h$. Under compressive strain, however, the CBM moves to lower potentials, causing a reduction in $U_e$, while the VBM shifts to higher (less negative) potentials, resulting in a decrease in $U_h$. As shown in Figure S8, the $Sb_2S_2O$ monolayer undergoes a transition from a direct to an indirect band gap under both biaxial compression and tensile strain. Under 6% compressive strain, the VBM shifts to the Γ-point while the CBM remains at the X-point. In contrast, under 6% tensile strain, the VBM relocates along the Γ-X direction, whereas the CBM still resides at the X-point. In contrast, both $Sb_2Se_2O$ and Janus $Sb_2SSeO$ monolayers preserve their indirect band gap character under biaxial strain. For both systems, the CBM consistently remains at the X-point under compressive and tensile conditions. At 6% compressive strain, the VBM is positioned along the X-$H_1$ direction. Under 6% tensile strain, however, the VBM shifts to the C-point in $Sb_2Se_2O$, while in the Janus $Sb_2SSeO$ monolayer it relocates to the $H_1$-point.

This wide tunability highlights the strong strain sensitivity of these 2D systems, which can be exploited to optimize their optoelectronic and photocatalytic performance. Accordingly, we evaluate their potential for large-scale photocatalytic water splitting by calculating the STH efficiencies using Eqs. (4)-(8). For the $Sb_2S_2O$ monolayer at pH = 7, the calculated efficiencies are $\eta_{abs}$= 9.74%, $\eta_{Cu}$= 39.12%, and $\eta_{STH}$= 3.81%. In contrast, the $Sb_2Se_2O$ monolayer is unable to drive overall water splitting at pH = 0, since its VBM lies above the $O_2/H_2O$ oxidation potential. However, at pH = 7, its calculated efficiencies are $\eta_{abs}$ = 22.65%, $\eta_{Cu}$ = 31.70%, and $\eta_{STH}$ = 7.18%. For the Janus $Sb_2SSeO$ monolayer, the calculated values are $\eta_{abs}$ = 19.91%, $\eta_{Cu}$= 32.60%, $\eta_{STH}$= 6.49%, and the corrected STH efficiency $\eta'_{STH}$ = 6.48%. These findings demonstrate that converting $Sb_2S_2O$ into its Janus counterpart enhances the solar-to-hydrogen efficiency by nearly 70%. Tables S3-S15 summarize the calculated $\eta_{abs}$, $\eta_{cu}$, $\eta_{STH}$, and $\eta'_{STH}$ for $Sb_2S_2O$, $Sb_2Se_2O$, and Janus $Sb_2SSeO$ monolayers for overall water splitting under various pH conditions and different levels of compressive and tensile biaxial strain. The maximum $\eta_{STH}$ of 5.83% was achieved for the $Sb_2S_2O$ monolayer under 4% compressive strain at pH = 7, representing a 55% enhancement compared to the unstrained structure. At pH = 7, the maximum solar-to-hydrogen efficiency is found to be 7.22% for the $Sb_2Se_2O$ monolayer under 4% tensile strain, and 7.78% for the Janus $Sb_2SSeO$ monolayer under 2% compressive strain. It is worth noting that the maximum STH efficiency achievable for a single photocatalyst is about 12%, whereas for heterojunction systems composed of two semiconductors, this value can be enhanced up to ~25% [50]. The STH efficiencies predicted for the $Sb_2X_2O$ and Janus $Sb_2SSeO$ monolayers are competitive with, or even exceed, those of many reported 2D photocatalysts, including covalent heptazine frameworks (~1.5–8.0%) [51], covalent triazine frameworks (0.30–3.70%) [51], and the $C_3N_2$ monolayer (4.54%) [52]. Higher STH values have also been reported for some other two-dimensional systems, including $In_2S_3$ (21.2%) and certain Janus monolayers such as SiAsP (11.87%) and SiSbP (20.15%) [53,54].

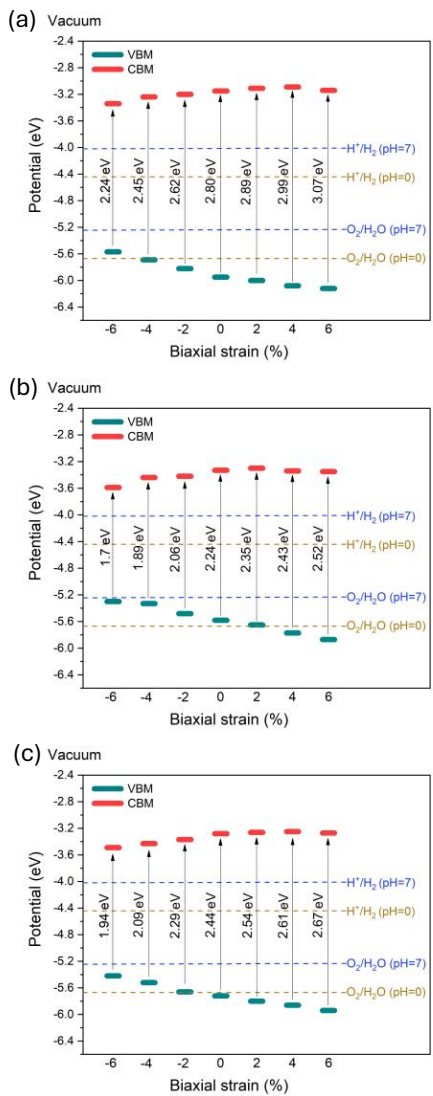

Figure 6. Band edge position of the VBM (green) and CBM (red) relative to the vacuum level for the (a) $Sb_2S_2O$, (b) $Sb_2Se_2O$, and (c) $Sb_2SSeO$ monolayers under different biaxial strains, calculated using the HSE06 functional. Band gap values are indicated in electron volts. Dashed brown and blue lines denote the water stability limits for hydrogen and oxygen evolution, respectively, at pH = 0 and pH = 7. The absolute potential of the standard hydrogen electrode (SHE), set at 4.44 eV at pH = 0 and 4.03 eV at pH = 7, is used as the reference.

**Thermodynamic Driving Forces for Water Splitting**

We subsequently conducted a comprehensive analysis of the mechanisms governing both the water oxidation and hydrogen reduction reactions on all the examined monolayers, employing the computational hydrogen electrode (CHE) model proposed by Nørskov et al. [55], as described in Supporting Information S11. For the HER and OER calculations, the optimized unit cell of each monolayer was used as the simulation supercell. The lateral lattice constants of $Sb_2X_2O$ ($a$ > 11 Å, $b$ > 8 Å) provide sufficient separation between periodic images of the adsorbed intermediates (H*, OH*, O*, OOH*), preventing spurious interactions. Convergence tests using slightly enlarged supercells confirmed that the adsorption energies and reaction free-energy profiles remain essentially unchanged. Hence, the conclusions regarding HER and OER activity are not sensitive to supercell size. As shown in Figure 7a, the water oxidation half-reaction proceeds through a four-electron (4e⁻) pathway, involving the sequential formation of key intermediates (*OH, *O, and *OOH or *O*OH). It is worth noting that two possible reaction routes were considered for the formation of the *OOH/*O*OH intermediates, namely, the single-site and dual-site mechanisms [51] (see Supporting Information S11). Moreover, the adsorption behavior of $H_2O$ molecules on the surfaces of these monolayers is discussed in detail in Supporting Information S12. For the OER process, all key intermediates preferentially adsorb on the Sb atoms that are bonded to oxygen, indicating that the water oxidation reaction predominantly occurs at these Sb active sites. Conversely, during the HER process (Figure 7b), hydrogen adsorption preferentially occurs on the oxygen atoms, proceeding via a two-electron (2e⁻) reaction pathway with H* as the key intermediate species. The corresponding Gibbs free energy (ΔG) profiles for the three monolayers are shown in Figures 7(c-h). In these plots, the black curves represent the reaction pathways without any applied potential, simulating the dark condition (U = 0 V). The red and purple curves correspond to the scenarios under applied external potentials ($U_e$ and $U_h$, respectively, as previously defined), which mimic the effect of photoexcited electrons and holes during illumination.

On the surfaces of these monolayers, the water oxidation reaction begins with the dissociation of the adsorbed $H_2O$ molecule to form an *OH intermediate, with free energy changes (ΔG) of 1.11 eV for $Sb_2S_2O$, 0.65 eV for $Sb_2Se_2O$, and 0.78 eV for $Sb_2SSeO$. The *OH species then loses a proton and an electron to produce *O, corresponding to ΔG values of 0.74, 1.18, and 0.80 eV, respectively. Subsequently, the *O intermediate reacts with another water molecule to form *OOH via the single-site mechanism, with ΔG values of 2.36 eV ($Sb_2S_2O$), 2.22 eV ($Sb_2Se_2O$), and 2.27 eV ($Sb_2SSeO$). This step represents the rate-determining step (RDS) with the highest energy barrier in the OER process. Finally, the *OOH species releases $O_2$, a proton, and an electron, accompanied by an exergonic energy release of 0.95 eV, 0.79 eV, and 0.60 eV, respectively. Alternatively, under the dual-site mechanism, the oxidation of *O to *O and *OH occurs with ΔG values of 1.21 eV ($Sb_2S_2O$), 1.31 eV ($Sb_2Se_2O$), and 1.55 eV ($Sb_2SSeO$), which remains the rate-determining step. In this mechanism, the subsequent conversion of *O*OH to $O_2$ is endergonic, requiring an additional 0.20 eV, 0.12 eV, and 0.12 eV for $Sb_2S_2O$, $Sb_2Se_2O$, and $Sb_2SSeO$ monolayers, respectively. The calculated ΔG values indicate that the dual-site mechanism is energetically more favorable than the single-site pathway, suggesting that cooperative interactions between neighboring active sites facilitate the OER process on these monolayer surfaces. As illustrated in Figure 7, under light illumination, photogenerated holes provide an external potential of 1.92, 1.55, and 1.75 V for the $Sb_2S_2O$, $Sb_2Se_2O$, and $Sb_2SSeO$ monolayers, respectively. However, in the single-site mechanism, the third reaction step remains energetically uphill, with ΔG values of 0.44, 0.67, and 0.52 eV for $Sb_2S_2O$, $Sb_2Se_2O$, and $Sb_2SSeO$, respectively. This indicates that the *O species cannot spontaneously convert into *OOH, and thus the OER process cannot proceed solely under illumination. Consequently, these materials are unlikely to achieve efficient OER without additional driving forces, such as an external bias or a co-catalyst. To initiate the OER through the single-site pathway, external potentials of at least 0.44, 0.67, and 0.52 eV are required for $Sb_2S_2O$, $Sb_2Se_2O$, and $Sb_2SSeO$, respectively, values that lie within the typical range reported for transition metal doped carbon (TM@C) or metal oxides (0.49–1.70 eV) [56]. Remarkably, in the dual-site mechanism, all reaction steps are thermodynamically downhill, indicating that each studied monolayer can effectively catalyze water oxidation under neutral conditions and light illumination.

Figure 7b illustrates the two-step process involved in the hydrogen reduction half-reaction. As shown in Figures 7d, f, and h, in the absence of light irradiation (U = 0 V), the first step involves the adsorption of a proton and an electron onto the $Sb_2S_2O$, $Sb_2Se_2O$, and $Sb_2SSeO$ monolayers, forming an *H intermediate with an unfavorable ΔG of 2.38, 2.01, and 1.92 eV, respectively. In the subsequent step, the adsorbed *H species combines with another proton and electron to generate a hydrogen molecule ($H_2$), an exergonic process that releases 1.55, 1.18, and 1.10 eV of energy for the same monolayers, respectively. Under the influence of the external potentials provided by photogenerated electrons (U = 0.88, 0.70, and 0.81 V for $Sb_2S_2O$, $Sb_2Se_2O$, and $Sb_2SSeO$ monolayers, respectively), the first hydrogen adsorption step remains endergonic but with a notably reduced energy barrier. The corresponding barriers decrease to 1.50, 1.30, and 1.11 eV at pH = 7 for the three monolayers, respectively. This reduction demonstrates that the Janus $Sb_2SSeO$ monolayer exhibits the most favorable kinetics for HER among the studied systems, although the overall reaction barrier remains relatively high. These energy barriers are either smaller than or comparable to those reported for typical photocatalysts, such as $Cd_6S_2$ monolayers [57], and $AgBiP_2Se_6$ monolayers [15], highlighting the competitive HER activity of the studied systems.

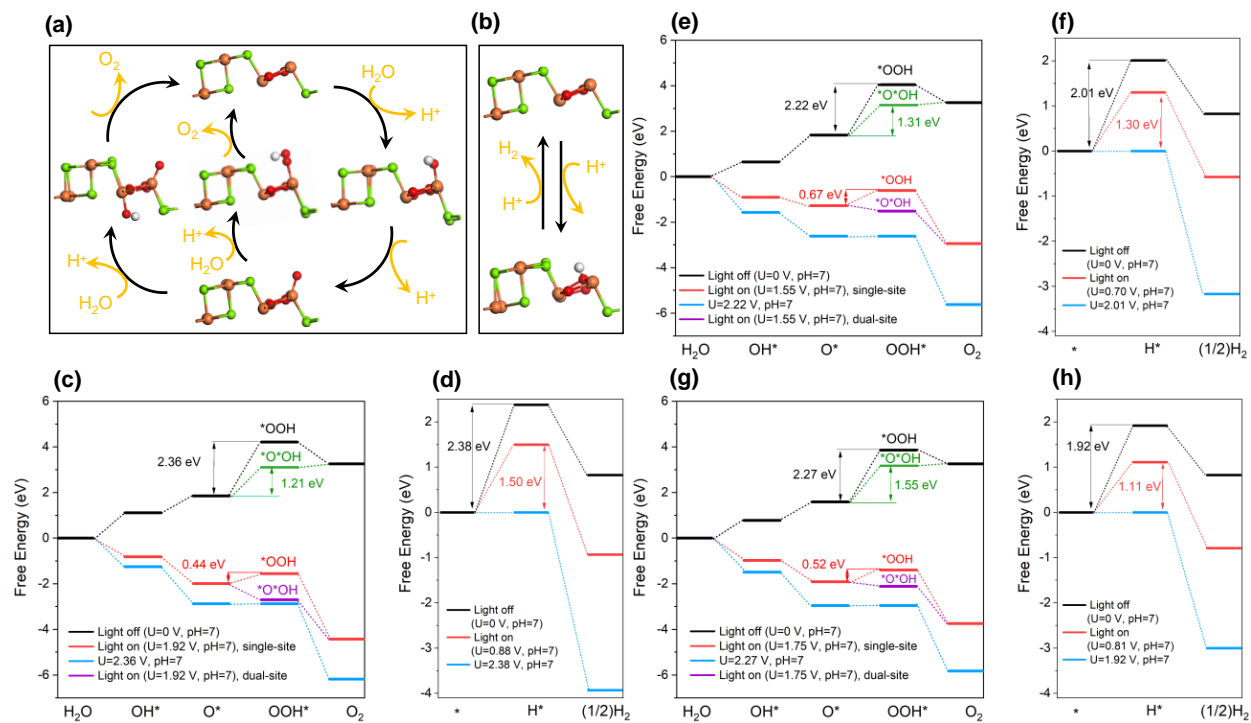

Figure 7. Photocatalytic pathways of (a) the oxygen evolution reaction (OER) and (b) the hydrogen evolution reaction (HER) on $Sb_2X_2O$ monolayers. Panels (c) and (d) present the Gibbs free energy diagrams of OER and HER, respectively, for the $Sb_2S_2O$ monolayer under both dark and illuminated conditions. Panels (e) and (f) show the corresponding diagrams for the $Sb_2Se_2O$ monolayer, while panels (g) and (h) depict the results for the $Sb_2SSeO$ monolayer. Atom color scheme: orange = Sb, green = S/Se, red = O, and white = H.

It is also important to note that efficient photocatalytic water splitting depends on several interconnected factors beyond the bandgap. A realistic assessment must consider proper band-edge alignment with HER and OER redox levels, strong light absorption, high carrier mobility, favorable reaction energetics, and good structural and chemical stability under operating conditions. These combined requirements help explain why practical solar-to-hydrogen efficiencies are often lower than theoretical predictions. In real applications, even promising materials may encounter challenges such as surface oxidation, photo-corrosion, defect formation, or limited carrier diffusion lengths. The variations in electronic and catalytic properties observed when substituting S/Se in $Sb_2X_2O$ originate mainly from differences in electronegativity, bonding strength, atomic radius, and induced lattice distortions, all of which directly influence band dispersion, effective masses, and intermediate adsorption energies. The reliability of our predictions is supported by the use of a consistent and well-validated computational protocol that includes phonon and AIMD stability checks, elastic and cleavage energy analyses, accurate HSE06 electronic structures, DFPT optical spectra, and detailed free-energy thermodynamics for all HER and OER steps. Practical application of these monolayers would require consideration of additional factors such as scalability, long-term stability in real aqueous environments, compatibility with co-catalysts, and integration into device architectures. Nevertheless, the combination of robust stability, favorable band alignment, strong optical response, efficient carrier transport, and promising HER/OER energetics suggests that $Sb_2X_2O$ and Janus $Sb_2SSeO$ monolayers remain strong candidates for future experimental validation and development toward next-generation photocatalytic systems.

**CONCLUSION**

Inspired by the recent experimental advance in the synthesis of layered $Sb_2S_2O$ [16], we conducted a comprehensive first-principles investigation aimed at the design and discovery of novel Sb-based oxychalcogenide monolayers. Our exploration of $Sb_2X_2O$ (X = S, Se, Te) and Janus $Sb_2SSeO$ monolayers considered structural, electronic, optoelectronic, and photocatalytic properties. Phonon dispersion, elastic constant analysis, and AIMD simulations confirm the thermodynamic and dynamical stability of $Sb_2S_2O$, $Sb_2Se_2O$, and $Sb_2SSeO$ monolayers, highlighting their stability for practical applications. The predicted cleavage energies, 0.36 J m$^{-2}$ for $Sb_2S_2O$ and 0.40 J m$^{-2}$ for $Sb_2Se_2O$, indicate weak interlayer coupling, confirming the feasibility of obtaining their suspended single-layer sheets through mechanical exfoliation. The excellent agreement between our HSE06-predicted bulk $Sb_2S_2O$ bandgap (2.00 eV) and the experimental value (2.04 eV) further validates the robustness and predictive accuracy of our computational framework. Electronic structure analysis reveals functional diversity within this monolayer family: $Sb_2S_2O$ is a direct-gap semiconductor (~2.80 eV), while $Sb_2Se_2O$ (~2.24 eV) and Janus $Sb_2SSeO$ (~2.44 eV) exhibit indirect band gaps. All monolayers display pronounced anisotropy in elasticity, carrier effective masses, and mobilities, highlighting their potential for direction-dependent optoelectronic applications. Moreover, the electronic and optical responses of each system are highly tunable under biaxial strain, with band gaps adjustable over wide ranges, offering a practical route for rational property modulation and device-specific design. A particularly notable finding is the Janus $Sb_2SSeO$ monolayer, which exhibits an intrinsic dipole; combined with favorable band alignment, strong optical absorption, efficient carrier transport, and relatively high dielectric constants, this establishes it as a highly promising photocatalyst for overall water splitting under neutral conditions. Overall, this study provides a solid theoretical foundation for the guided design and experimental realization of $Sb_2X_2O$ and Janus $Sb_2SSeO$ nanosheets, demonstrating anisotropy, strain tunability, and strong photocatalytic performance, highlighting their potential in next-generation direction-dependent optoelectronic devices and sustainable energy-conversion technologies.

## ASSOCIATED CONTENT

### Supporting Information

Structure of $Sb_2X_2O$ bilayer; crystal structure of $Sb_2S_2O$ bulk phase; phonon dispersion of single-layer $Sb_2Te_2O$; structure of the Janus $Sb_2SSeO$ monolayer; electronic properties of the $Sb_2S_2O$ bulk structure; the partial orbital density of states $Sb_2S_2O$, $Sb_2Se_2O$, and $Sb_2SSeO$ monolayers; thermodynamic oxidation and reduction potentials ($\Phi^{ox}$, $\Phi^{re}$) for the $Sb_2X_2O$ and $Sb_2SSeO$ monolayers, evaluated relative to the water redox levels; dielectric constant and absorption coefficient; electronic band structure under different biaxial strains; solar-to-hydrogen (STH) conversion efficiency under different biaxial strains; free energy difference (ΔG) for photocatalytic water splitting; $H_2O$ adsorption, and the structures of $Sb_2X_2O$ and $Sb_2SSeO$ monolayers in POSCAR format.

## AUTHOR INFORMATION

### Corresponding Author

Masoud Shahrokhi − Department of Physics, Faculty of Science, University of Ostrava, 30. dubna 22, 701 03 Ostrava, Czech Republic; orcid.org/0000-0003-3656-6551

Email: shahrokhimasoud37@gmail.com

### Author

Bohayra Mortazavi − Institute of Photonics, Department of Mathematics and Physics, Leibniz Universität Hannover, Welfengarten 1A, Hannover 30167, Germany; Cluster of Excellence PhoenixD, Leibniz Universität Hannover, Welfengarten 1A, Hannover 30167, Germany; orcid.org/0000-0003-3031-5057

## ACKNOWLEDGMENTS

M.S. acknowledges the financial support of the European Union under the LERCO project number CZ.10.03.01/00/22_003/0000003 via the Operational Programme Just Transition and the Ministry of Education, Youth and Sports of the Czech Republic through the e-INFRA CZ (ID:90254). B.M. appreciates the funding by the Deutsche Forschungsgemeinschaft, Germany (DFG, German Research Foundation) under Germany's Excellence Strategy within the Cluster of Excellence PhoenixD (EXC 2122, Project ID 390833453).